\documentclass[a4paper,11pt]{article}
\usepackage[T1]{fontenc}
\usepackage[utf8]{inputenc}
\usepackage{newtxtext}
\usepackage{amsthm}
\usepackage{newtxmath}
\usepackage{amsmath}
\usepackage{graphicx}
\usepackage[margin=25mm]{geometry}
\usepackage[numbers,compress]{natbib}
\usepackage[font=small,labelfont=bf]{caption}
\usepackage{authblk}

\usepackage{xcolor}
\usepackage[colorlinks=true,linkcolor=blue!60!black,citecolor=blue!60!black,urlcolor=blue!60!black]{hyperref}

\newcommand\acknowledgments{\section*{Acknowledgments}}

\title{Accelerating experimental spintronics research through context provisioning for LLM agents}

\author[1,*]{Aditya Kumar}
\author[1]{Kilian Leutner}
\author[1,*]{Mathias Kl\"aui}
\affil[1]{Institute of Physics, Johannes Gutenberg University Mainz, Staudingerweg 7, 55128 Mainz, Germany}
\affil[*]{Corresponding authors: \href{mailto:kuaditya@uni-mainz.de}{kuaditya@uni-mainz.de}, \href{mailto:klaeui@uni-mainz.de}{klaeui@uni-mainz.de}}
\date{}

\begin{document}
\maketitle

\begin{abstract}
Experimental spintronics combines solid-state physics, thin-film deposition, nanofabrication, and spin-sensitive characterization. Sample fabrication and characterization rely on customized apparatus. Automated deposition and measurement systems exist, but each experimental run remains resource-intensive and time-consuming, so raising throughput by running more experiments scales these demands proportionally. We therefore pursue a lower-cost lever, the information gained per experiment, by targeting the planning and analysis stages and giving agents based on large language models access to three sources of context: the scientific literature, internal laboratory documentation, and fast, routine characterization instruments, each exposed through the Model Context Protocol. Two on-premise retrieval systems, lit-RAG and wiki-RAG, provide the literature and documentation context, including figures. On a 10,000-document physics corpus, embedding figure captions retrieves the correct figure among the top ten results for 92\% of test questions (recall@10 = 0.92), compared with 0.04--0.05 for vision--language image embeddings. Because domain terms recur across many documents in this corpus, fusing lexical with dense search lowers recall, whereas on the laboratory documentation dense retrieval reaches a page recall@10 of 0.98 and fusion does not lower it. For instruments, we describe a control architecture, demonstrated on Kerr microscopy, that lets agents operate laboratory hardware within declared physical limits.
\end{abstract}

The manipulation of spin degrees of freedom in solid-state systems underpins modern information technology, enabling non-volatile data storage, high-sensitivity magnetic sensors, and energy-efficient spin-transfer torque magnetic random-access memory~\cite{dieny2020opportunities,bhatti2017spintronics}. Artificial intelligence (AI) requires massive computing power, and spintronics can offer low-energy solutions for this~\cite{bibes2026applied}. Developing next-generation spintronics requires exploring multi-dimensional parameter spaces, executing intricate nanofabrication recipes, and conducting precision measurements on often customized or home-built apparatus. Magnetic skyrmions and domain walls are actively investigated for physical computing~\cite{fert2017skyrmions,raab2022brownian,beneke2024gesture,kumar2022domainwall} and for multi-turn rotation sensing~\cite{borie2017multiturn,leutner2023skyrmion}. Domain-wall turn counters are patented~\cite{mattheis2014revolution} and commercial~\cite{novotechnik2026novoturn,adi2026admt4000}, and a skyrmion turn counter is patented~\cite{klaeui2025skyrmioncounter}, but most of these concepts are still far from applications. An opportunity is to use AI to support the experimental workflow and accelerate research. Agents based on large language models (LLMs) can assist in hypothesis generation, planning, and data analysis on existing laboratory infrastructure.

\section{AI for experimental magnetism: context for planning and analysis}

Autonomous laboratories have demonstrated high throughput in standardized chemical synthesis and fluidic materials discovery~\cite{burger2020mobile,szymanski2023autonomous,boiko2023autonomous,dai2024autonomous}. However, academic experimental condensed-matter and spintronics laboratories rely on customized apparatus with limited sample throughput, where physical robotics cannot be readily scaled. We therefore target the planning and analysis stages, which determine the information yield per experiment and do not require robotics. LLM agents have demonstrated scientific reasoning capabilities, from planning and executing chemical experiments~\cite{boiko2023autonomous,bran2024chemcrow,muller2026agenteoptima} to generating and ranking research hypotheses~\cite{gottweis2026coscientist}. We supply an agent with three context sources: (1) the \emph{laboratory context} (instrument configurations, handling protocols, fabrication recipes, and other institutional know-how recorded in the group wiki), (2) the \emph{external literature context} (peer-reviewed publications and theses), and (3) the \emph{experiment context} (live instrument readings and actuation). Each source is exposed as an independent tool server via the open Model Context Protocol (MCP)~\cite{anthropic2024mcp} and queried by LLMs hosted on campus, so that no internal data leaves the institution. Retrieval-augmented generation (RAG)~\cite{lewis2020rag} grounds agent outputs in sources that can be checked, which reduces hallucination in technical domains~\cite{asai2026synthesizing,chiang2025llamp}.

\section{Provenance-preserving retrieval over a group literature corpus}

Synthesizing literature is essential for experimental design. New arXiv submissions grew from about 105,000 in 2015 to about 284,000 in 2025~\cite{arxiv2026stats}, which makes it difficult for researchers to stay up to date. Systems like OpenScholar~\cite{asai2026synthesizing} index open-access papers in coarse blocks, omitting licensed articles and figures, although figures hold the key results of most experimental papers. Our system, lit-RAG, indexes the full literature corpus individually collected by the group, including licensed articles ($\approx$10,000 documents). Documents are parsed with Docling~\cite{docling2024} and GROBID~\cite{lopez2009grobid} into a typed block tree preserving bounding-box provenance and citation links. Passages (up to 510 tokens) are embedded with BGE-M3~\cite{chen2024bge} (1024 dimensions) in PostgreSQL with pgvector (hierarchical navigable small-world index). Figures are indexed as text units of label and caption. Via MCP, agents run dense searches filtered by metadata and traverse citation links. Each benchmark has 100 questions generated by an LLM from random passages or figures, filtered by a second LLM, and checked by the authors to be specific, answerable, and not verbatim, as in LitSearch~\cite{ajith2024litsearch}. Each question thus has one correct answer, the passage or figure it was generated from. We report recall@$k$ (R@$k$), the fraction of questions for which this item is among the top $k$ results; at paper (page) level, any passage of the source document (wiki page) counts.

\begin{figure}[t]
  \centering
  \includegraphics[width=\textwidth]{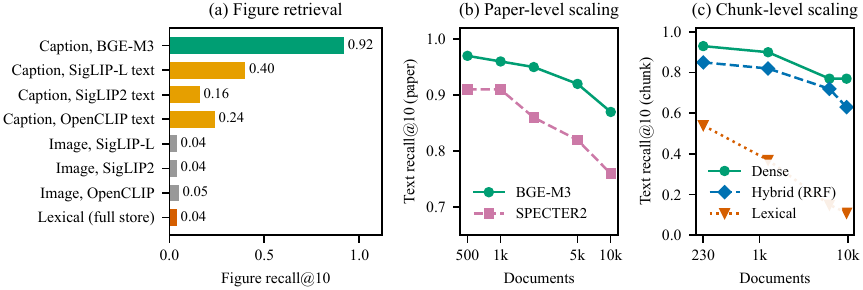}
  \caption{\textbf{Retrieval performance of lit-RAG}. (a) Figure recall@10 for BGE-M3 captions vs.\ vision--language text/image encoders and lexical search. (b) Paper-level recall@10 vs.\ corpus size for BGE-M3 and SPECTER2. (c) Chunk-level recall@10 for dense, hybrid (reciprocal rank fusion, RRF), and lexical retrieval with BGE-M3.}
  \label{fig:results}
\end{figure}

\subsection{Figure retrieval: caption text versus image embeddings}

Scientific figures are multi-panel compositions that cannot be interpreted without the measurement conditions and parameters defined in their captions, which makes the raw figures incomplete for accurate retrieval. On a pool of 63,825 figures, we compared (A) caption text encoded with BGE-M3, (B) figure images with the image encoders of three vision--language models, SigLIP-large~\cite{zhai2023siglip}, SigLIP2-so400m~\cite{tschannen2025siglip2}, and OpenCLIP ViT-L/14~\cite{radford2021clip,ilharco2021openclip}, and (C) caption text with the text encoders of the same models. With BGE-M3 caption embeddings, the correct figure is among the top 10 results for 92 of the 100 questions (R@10 = 0.92) and ranked first for 85 (R@1 = 0.85). All image encoders reach R@10 between 0.04 and 0.05, comparable to lexical search (0.04; Fig.~\ref{fig:results}a). The difference is significant in a paired McNemar test~\cite{mcnemar1947note} ($p < 10^{-25}$ for each image encoder), and for every model the text encoder (C) outperforms the image encoder (B).
\subsection{Encoder choice, corpus scaling, and retrieval modes}

On 230 documents (510-token chunks), text chunk R@10 is 0.56 and 0.75 for the citation-trained encoders SciNCL~\cite{ostendorff2022neighborhood} and SPECTER2~\cite{singh2022scirepeval}, versus 0.93 for BGE-M3 ($p < 10^{-4}$ vs.\ SPECTER2, McNemar test). As the corpus grows from 500 to 10,019 documents, BGE-M3 paper R@10 decreases from 0.97 to 0.87 (Fig.~\ref{fig:results}b). From 230 to 9,642 documents, BGE-M3 chunk R@10 falls from 0.93 to 0.77, whereas lexical chunk R@10 (PostgreSQL full-text ranking, without inverse document frequency weighting) falls from 0.54 to 0.11 (Fig.~\ref{fig:results}c); frequent terms discriminate poorly (``Pt'' occurs in 4,002 and ``skyrmion'' in 2,764 of 9,965 documents). With BGE-M3, hybrid retrieval, which fuses the dense and lexical rankings by reciprocal rank fusion (RRF)~\cite{cormack2009reciprocal}, scored below dense retrieval at every measured corpus size (chunk R@10 0.85 vs.\ 0.93 at 230 documents, 0.63 vs.\ 0.77 at 9,642), so we use dense retrieval alone. We attribute this to the recurrent domain terms, which our lexical ranking does not down-weight. Lexical search also lags behind dense retrieval on the LitSearch benchmark of scientific-literature questions, where the recall@5 of the standard lexical ranking function BM25~\cite{robertson2009bm25} is 24.8 percentage points below that of the best dense retriever~\cite{ajith2024litsearch}.

\section{Retrieval over internal laboratory documentation}

wiki-RAG exposes the group wiki (749 pages indexed), which records setup conventions, measurement artifacts, and institutional know-how, to agents through an on-premise MCP server. On 100 questions built as for lit-RAG, page-level R@10 is 0.98 for dense BGE-M3 retrieval and 0.92 for lexical. Unlike in the literature corpus, lexical search remains strong here and fusing the two does not lower recall, so wiki-RAG serves hybrid retrieval; the value of fusion depends on the corpus and the lexical scorer.

\section{Agentic instrument control and staged integration}

Physical execution introduces operational hazards: unconstrained LLM actuation risks equipment or sample damage. Our framework therefore separates high-level model reasoning from low-level execution via a virtual-instrument layer, a centralized orchestration layer, and a security layer. Each instrument declares bounded parameter ranges in one declarative, human-readable specification file, from which a graphical user interface, an MCP server, and a command-line interface are generated. Actions are converted into a unified command type and validated against the declared limits, caller roles checked by the security layer, and a software emergency stop before execution, with every event recorded in an append-only audit log. This enables staged autonomy: beginning with read-only monitoring, advancing to supervised observation, sandboxed analysis, and bounded execution, as demonstrated in our agentic Kerr microscope deployment~\cite{leutner2026agentkerr}. 

\section{Conclusion and outlook}

Our measurements show that caption embedding outperforms image embedding for figure retrieval (0.92 vs.\ $\le 0.05$ R@10), that for question-based search BGE-M3 outperforms citation-trained encoders, and that rank fusion with lexical search lowered dense recall on the literature corpus but not on the wiki. We continue to work in these three directions: (1) improving retrieval in lit-RAG in the continuously growing corpus with domain fine-tuning and cross-encoder reranking; (2) using wiki-RAG questions that fail or are abstained on to locate gaps in the documentation of processes and setups, and LLMs to help close them; and (3) expanding instrument control beyond Kerr microscopy to cryogenic magnetotransport setups.

\acknowledgments

This work is funded by the Deutsche Forschungsgemeinschaft (DFG), TRR 173 Spin+X (Projects A01, A03, A11, A12, B02, B15) No.\ 268565370, DFG Project No.\ 358671374, and the Research Council of Norway (QuSpin No.\ 262633).

\bibliographystyle{JHEP}
\bibliography{references}

\end{document}